\documentclass[aps,prb,showpacs,amsmath,amssymb,reprint,superscriptaddress,floatfix]{revtex4-1}

\usepackage[colorlinks=true,citecolor=blue]{hyperref} 
\usepackage{graphicx}
\graphicspath{{figures2/}}
\usepackage{xcolor}
\usepackage{ulem}

\begin{document}

\title{Controlling magnetism through Ising superconductivity in magnetic van der Waals heterostructures}

\author{Faluke Aikebaier}
\email[]{faluke.aikebaier@gmail.com}
\affiliation{Department of Applied Physics, Aalto University, 00076, Espoo, Finland}
\affiliation{Department of Physics and Nanoscience Centre, University of Jyv\"askyl\"a, P.O. Box 35, fi-40014 University of Jyväskylä, Finland}
\affiliation{Computational Physics Laboratory, Physics Unit, Faculty of Engineering and Natural Sciences, Tampere University, FI-33014 Tampere, Finland}

\author{Tero T. Heikkilä}
\affiliation{Department of Physics and Nanoscience Centre, University of Jyv\"askyl\"a, P.O. Box 35, fi-40014 University of Jyväskylä, Finland}

\author{J. L. Lado}
\affiliation{Department of Applied Physics, Aalto University, 00076, Espoo, Finland}

\begin{abstract}
Van der Waals heterostructures have risen as a tunable platform to combine different electronic orders, due to the flexibility in stacking different materials with competing symmetry broken states. Among them, van der Waals ferromagnets such as $\rm{CrI}_3$, CrBr$_3$ or CrCl$_3$ and superconductors as $\rm{NbSe}_2$ provide a natural platform to engineer novel phenomena at ferromagnet-superconductor interfaces. In particular, NbSe$_2$ is well known for hosting strong spin-orbit coupling effects that influence the properties of the superconducting state. Here we put forward a ferromagnet/$\rm{NbSe}_2$/ferromagnet heterostructure where the interplay between Ising superconductivity in $\rm{NbSe}_2$ and magnetism controls the magnetic alignment of the heterostructure. In particular, we show that the interplay between spin-orbit coupling and superconductivity
provides a new mechanism  to control magnetic ordering in van der Waals materials. We show that this coupling allows creating heterostructures featuring a magnetic phase transition from in-plane to out-of-plane associated to the onset of superconductivity. Our results show how hybrid van der Waals ferromagnet/superconductor heterostructure can be used as a tunable materials platform for superconducting spin-orbitronics.

\end{abstract}

\maketitle

\section{Introduction}

Van der Waals (vdW) heterostructures have become one of the paradigmatic
platforms to engineer controllable quantum materials~\cite{novoselov20162d,Geim2013}. This
flexibility stems from the possibility of combining in a single structure
a variety of competing electronic orders\cite{liu2016van}, including
semimetals, insulators, semiconductors, superconductors, and ferromagnets
among others\cite{novoselov20162d,Geim2013,liu2016van}. In particular, vdW heterostructures provide a natural platform to engineer novel phenomena at interfaces of antagonist orders, including superconductivity~\cite{PhysRevLett.86.4382,saito2016highly,qiu2021recent} and ferromagnetism~\cite{shabbir2018long,gong2017discovery,huang2017layer,zhang2019direct,ghazaryan2018magnon},
that can potentially lead to a whole new family of superconducting-spintronic devices\cite{Kezilebieke2020,Kezilebieke2021,2021arXiv210101327K,2020arXiv201109760K}.

Monolayer transition metal dichalcogenides (TMD)\cite{manzeli20172d}
provide a rich playground to exploit spin-orbit driven
phenomena due to their large Ising spin-orbit coupling (SOC) effects~\cite{PhysRevB.84.153402}. 
Their intrinsic spin-orbit couplings lead to a momentum-dependent spin splitting 
generating robust spin-momentum locking,~\cite{PhysRevB.84.153402,PhysRevLett.108.196802}
a highly attractive feature for spintronics and valleytronics~\cite{PhysRevLett.108.196802,kormanyos2014spin,Soriano2020,PhysRevLett.122.086401,PhysRevB.101.045408,PhysRevB.100.085128}. 
In particular, NbSe$_2$ develops a so-called 
Ising superconducting state~\cite{Xu2014,xi2016ising}, as a direct consequence of the interplay of spin-singlet
superconductivity and Ising spin-orbit coupling~\cite{PhysRevLett.108.196802}, leading
to a dramatic enhancement of the in-plane critical field~\cite{lu2015evidence,xi2016ising,saito2016superconductivity,youn2012role,sergio2018tuning}. 
As a result, NbSe$_2$
provides a suggestive platform to explore the novel interplay between
magnetism, superconductivity, and Ising spin-orbit coupling\cite{2021arXiv210101327K,Rahimi2017}.

\begin{figure}[t]
\includegraphics[width=1.0\linewidth]{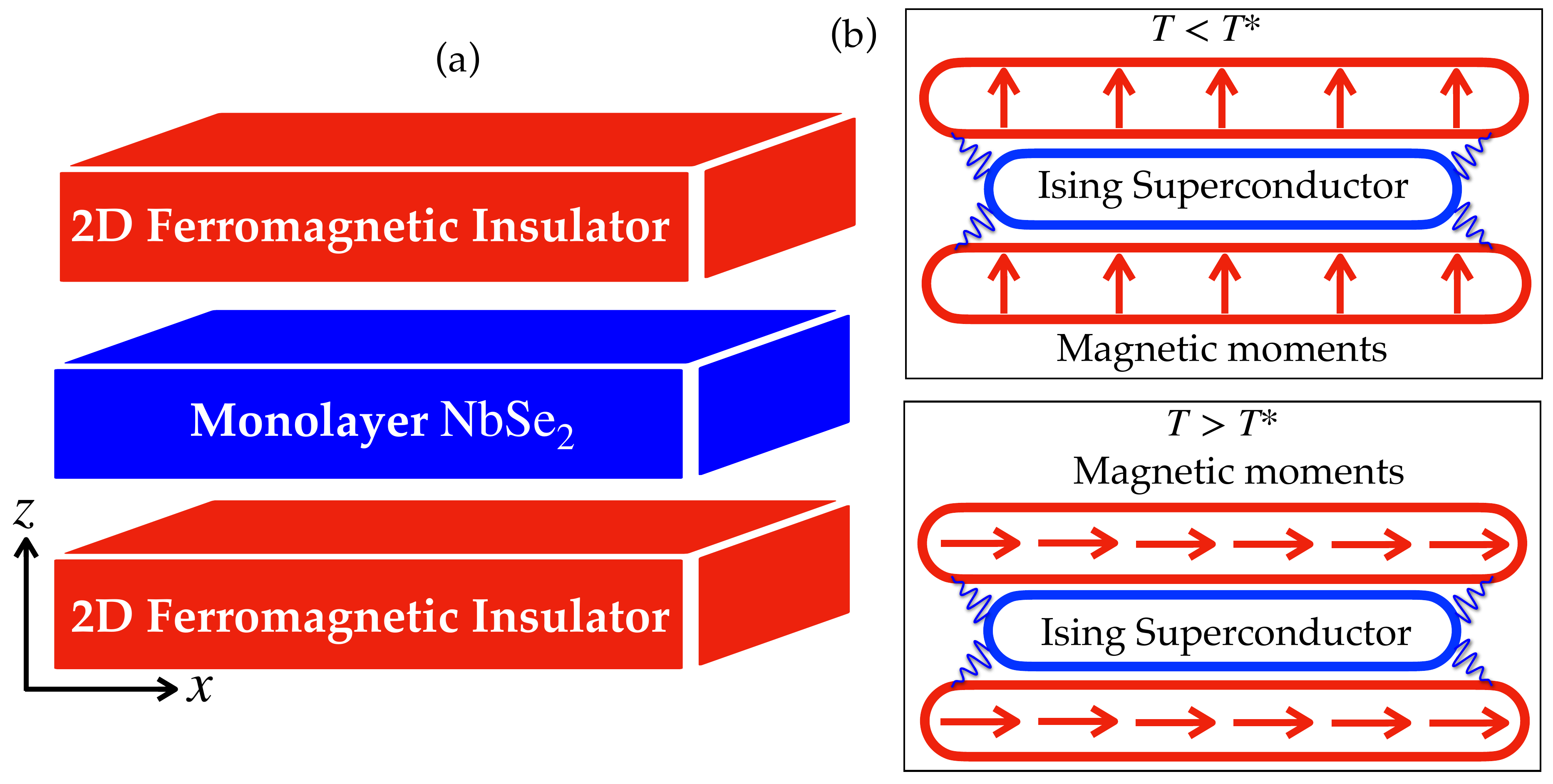}%
\caption{\label{fig:structure} (a) Schematic structure of the vdW heterostructure considered in this paper, with ferromagnet/${\rm{NbSe}}_2$/ferromagnet sequence. Here $x$ and $z$ axes are the Cartesian axes. (b) Schematic illustration of the coupling between ferromagnetic layers through an Ising superconductor. For $T<T^*$ the ferromagnetic coupling is out-of-plane, and for $T>T^*$ the ferromagnetic coupling is in-plane, where $T^*$ is a temperature below the superconducting critical temperature $T_c$. }
\end{figure}

Here we put forward a magnet/superconductor van der Waals heterostructure as a controllable system where
a magnetic transition is driven purely by Ising superconductivity.
Specifically, we show that a monolayer-ferromagnet/NbSe$_2$/monolayer-ferromagnet
[Fig.~\ref{fig:structure}(a)] leads to 
an artificial system displaying magnetic transitions triggered by Ising superconductivity. 
We demonstrate how the interplay between the Ising superconductivity and the ferromagnetic
proximity effect
controls the magnetic alignment of the heterostructure. In particular, we show that the Ising SOC keeps the magnetic alignment of the heterostructure in-plane in the normal state, and drives a transition from in-plane to out-of-plane magnetic alignment  in the superconducting state [Fig.~\ref{fig:structure}(b)]. Our results put forward Ising superconductivity as a potential knob to control
magnetism in artificial heterostructures, leading to
a promising minimal building block for
van-der-Waals-based superconducting
spintronics.

The manuscript is organized as follows. First, in section \ref{sec:model} we present the effective model used to capture the
physics of the ferromagnet/NbSe$_2$/ferromagnet heterostructures. In section \ref{sec:ising} we discuss the interplay between Ising spin-orbit coupling, exchange coupling, and superconductivity. In section \ref{sec:main} we show
the emergence of magnetic switching
driven by Ising superconductivity. Finally, in section \ref{sec:con} we summarize our results.

\section{Model}
\label{sec:model}

In the following we consider a multilayer structure consisting of a monolayer ${\rm{NbSe}}_2$ sandwiched between two ferromagnets, as shown in Fig.~\ref{fig:structure}(a). The proximity effect between the ferromagnets and the superconducting layer induces an exchange field in the ${\rm{NbSe}}_2$. The direction of the induced field depends on the magnetization directions in the two ferromagnets. Overall, the net Hamiltonian of the system is

\begin{equation}\label{eq:TotalHamiltonian}
    H = H_0 + H_{SC} + H_{J},
\end{equation}
where $H_0$ is the single-particle Hamiltonian of NbSe$_2$ including spin-orbit coupling,
$H_{SC}$ accounts for the superconducting state and
$H_{J}$ includes the effect of the ferromagnetic leads obtained
by integrating out the CrI$_3$ degrees of freedom.

\begin{figure}[t]
\includegraphics[width=\linewidth]{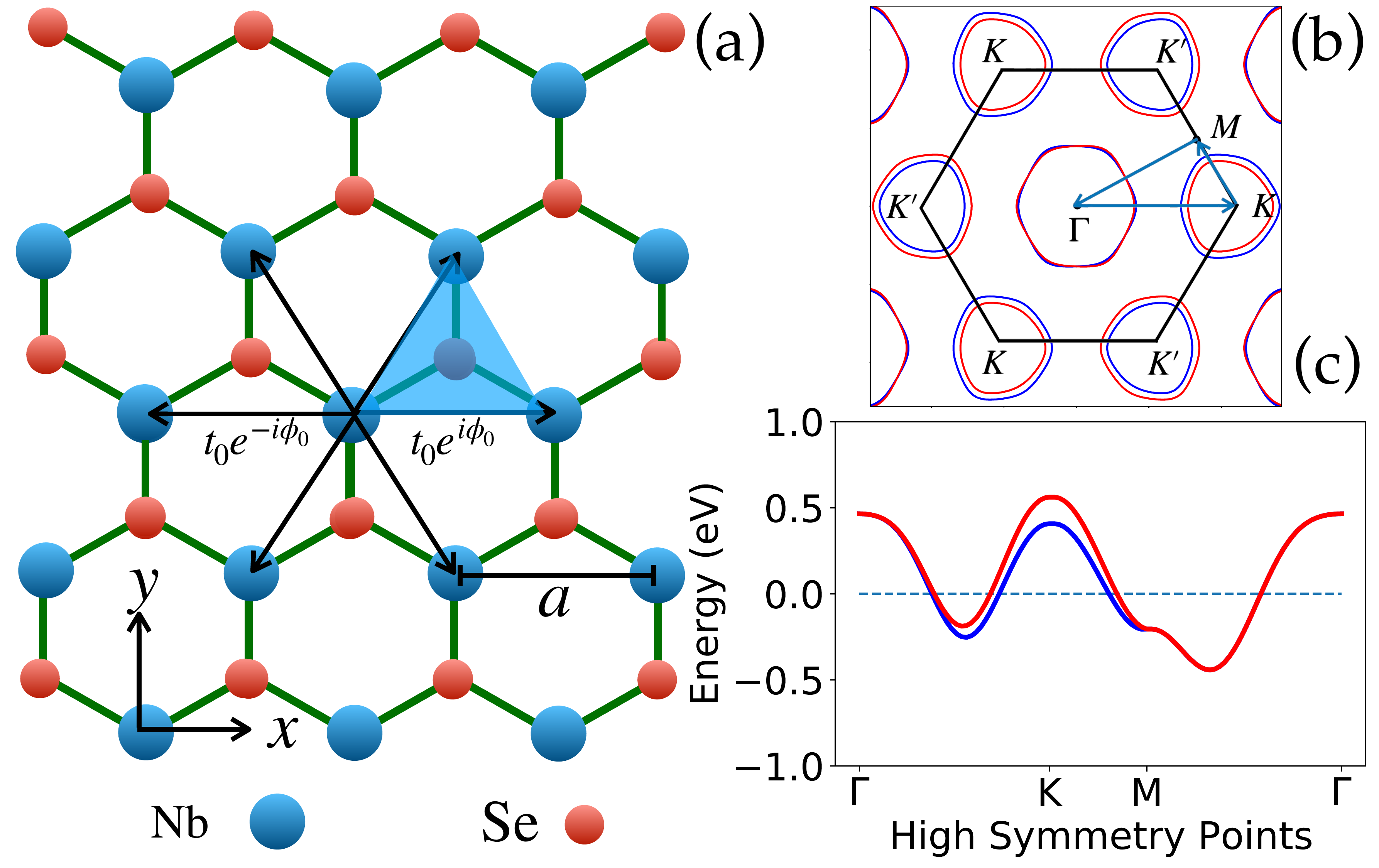}%
\caption{\label{fig:tight_binding} (a) Top view of the crystal (hexagonal) structure of monolayer $\rm{NbSe}_2$. Due to the sublattice imbalance, it can be effectively treated as a triangular lattice with one $d$ orbital per site. The shaded area is the unit cell of the triangular lattice and the black arrows denote the unit vectors connecting nearest neighbours with complex hopping parameters. The lattice constant $a$ is the distance between Nb atoms. The spin quantization axis is in the out-of-plane direction. (b) Brillouin zone of monolayer $\rm{NbSe}_2$ together with the energy bands at the Fermi level. (c) Band structure of the effective single-band model around the Fermi level of monolayer $\rm{NbSe}_2$. The band structure is along the high symmetry points denoted as the blue arrows in (b). The red and the blue curves represent spin up and spin down components, respectively.}
\end{figure}

Let us first introduce the single-particle model $H_0$ for the ${\rm{NbSe}}_2$ layer\cite{PhysRevB.92.205108,Liebhaber2019}. To study the superconducting properties, only the low-energy part of the band structure is sufficient for consideration. The crystal structure of the monolayer $\rm{NbSe}_2$ has the $D_{3h}$ point-group symmetry, strongly
constraining  its lowest energy model.
Since a single band is located at the Fermi energy~\cite{lebegue2009electronic}, a 
Wannier Hamiltonian with a single orbital
in the triangular lattice can be built [Fig.~\ref{fig:tight_binding}(a)]. 
The Wannier low energy model for NbSe$_2$ takes the form
\begin{equation}\label{eq:OneBandTighBindingHamiltonian}
H_0=
\sum_{\alpha \beta} t^{ss'}_{\alpha \beta}
c^\dagger_{\alpha,s} c_{\beta,s'},
\end{equation}
where $t^{ss'}_{\alpha \beta}$ are the spin-dependent hopping 
amplitudes\footnote{The values of the tight-binding parameters are given in
appendix.~\ref{sec:EffectiveTBModel}},
and $c^\dagger_{i,s}$ is the creation operator of a Wannier orbital centered
at each Nb site.
This model captures the low-energy band structure of the monolayer $\rm{NbSe}_2$. The SOC is included by a Kane-Mele form~\cite{kane2005quantum} as a complex spin and bond dependent first neighbor
complex hopping
parameterized by a phase $\phi_i$. We include
up the sixth neighbor hopping between Wannier orbitals. 
The resulting electronic structure is shown in
Fig.~\ref{fig:tight_binding}(c). 

The magnetic proximity effect with the top and bottom ferromagnets is included
in the electronic structure of NbSe$_2$ as~\cite{khusainov1996indirect,izyumov2002competition}

\begin{equation}\label{eq:ExchnageHamiltonian}
    H_{J} = 
    \sum_k \boldsymbol{J}\cdot\boldsymbol{\sigma}_{ss'} 
    c_{\boldsymbol{k}s}^{\dagger}c_{\boldsymbol{k}s'}.
\end{equation}

The superconducting term in the Hamiltonian is a conventional s-wave
superconducting order of the form
\begin{equation}\label{eq:HamiltonianSCi}
H_{SC} = 
\sum_{\boldsymbol{k},\uparrow,\downarrow}\left(\Delta c_{\boldsymbol{k},\uparrow}^{\dagger}c_{-\boldsymbol{k},\downarrow}^{\dagger}+\Delta^* c_{-\boldsymbol{k},\downarrow}c_{\boldsymbol{k},\uparrow} \right)+K,
\end{equation}
where $c_{\boldsymbol{k},s}^{(\dagger)}$ is the annihilation (creation) operator with spin $s$ at momentum $\boldsymbol{k}$, $\boldsymbol{\sigma}$ is the vector of Pauli spin matrices and $\boldsymbol{J}=(J_x,J_y,J_z)$ is the vector of induced exchange field in the ${\rm{NbSe}_2}$ layers. It can be expressed in the spherical coordinates as
$
J_x=J\sin\theta\cos\varphi$,
$
J_y=J\sin\theta\sin\varphi$,
and
$
J_z=J\cos\theta,
$ 
where $J$ is the strength of the induced exchange field in the superconductor layer, and $\theta$ and $\varphi$ are the related polar and azimuthal angles, respectively.

The superconducting pair potential is determined self-consistently as
$
\Delta=g \int \left\langle c_{-\boldsymbol{k}\uparrow}c_{\boldsymbol{k}\downarrow} \right\rangle d^2 \mathbf{k},
$ 
where $g$ is the interaction potential. The last term in Eq.~\eqref{eq:HamiltonianSCi} then can be written as
\begin{equation}\label{eq:FreeEnergyDensityConstant}
K=g\int \left\langle c_{\boldsymbol{k}\uparrow}^{\dagger}c_{-\boldsymbol{k}\downarrow}^{\dagger} \right\rangle\Big\langle c_{-\boldsymbol{k}\downarrow}c_{\boldsymbol{k}\uparrow} \Big\rangle d^2 \mathbf{k}=\frac{\vert\Delta\vert^2}{g}. 
\end{equation}

The contribution of superconductivity to the energetics of the system can be obtained by considering the free energy density. In general, the free energy density can be derived from the partition function~\cite{landau2013statistical} $Z=e^{-K/k_BT}\prod_{\boldsymbol{k},n,\sigma}\left(1+e^{-\epsilon_{\boldsymbol{k},n,\sigma}/k_BT} \right)$, 
where $K$ is the constant term in the Hamiltonian in Eq.~\eqref{eq:HamiltonianSCi}, $\epsilon_{\boldsymbol{k},n,\sigma}$ are the eigenvalues (with spin $\sigma$) of the Hamiltonian of the system under consideration, $T$ is the temperature and $k_B$ is the Boltzmann constant. We set $k_B=1$ hereafter. The band index $n$ runs only over the empty states. The free energy density is defined as $f=-T\log Z$. The detailed form of the free energy density can be written as
\begin{equation}\label{eq:FreeEnergyDensityGeneral}
f=\frac{\vert\Delta\vert^2}{g} -T\sum_{\boldsymbol{k},n}\log \left[2+2\cosh\left(\frac{\epsilon_{\boldsymbol{k},n}}{T} \right)\right].
\end{equation}
In the following, we concentrate on the free energy density $f_s$ of the superconducting state in the presence of exchange field compared to the free energy density $f_n$ in the normal state in the absence of the exchange field  $f_{sn}=f_s({\bf J})-f_n(J=0)$. 

\section{Impact of exchange field on Ising Superconductivity }
\label{sec:ising}

\begin{figure}[t]
\includegraphics[width=1.0\linewidth]{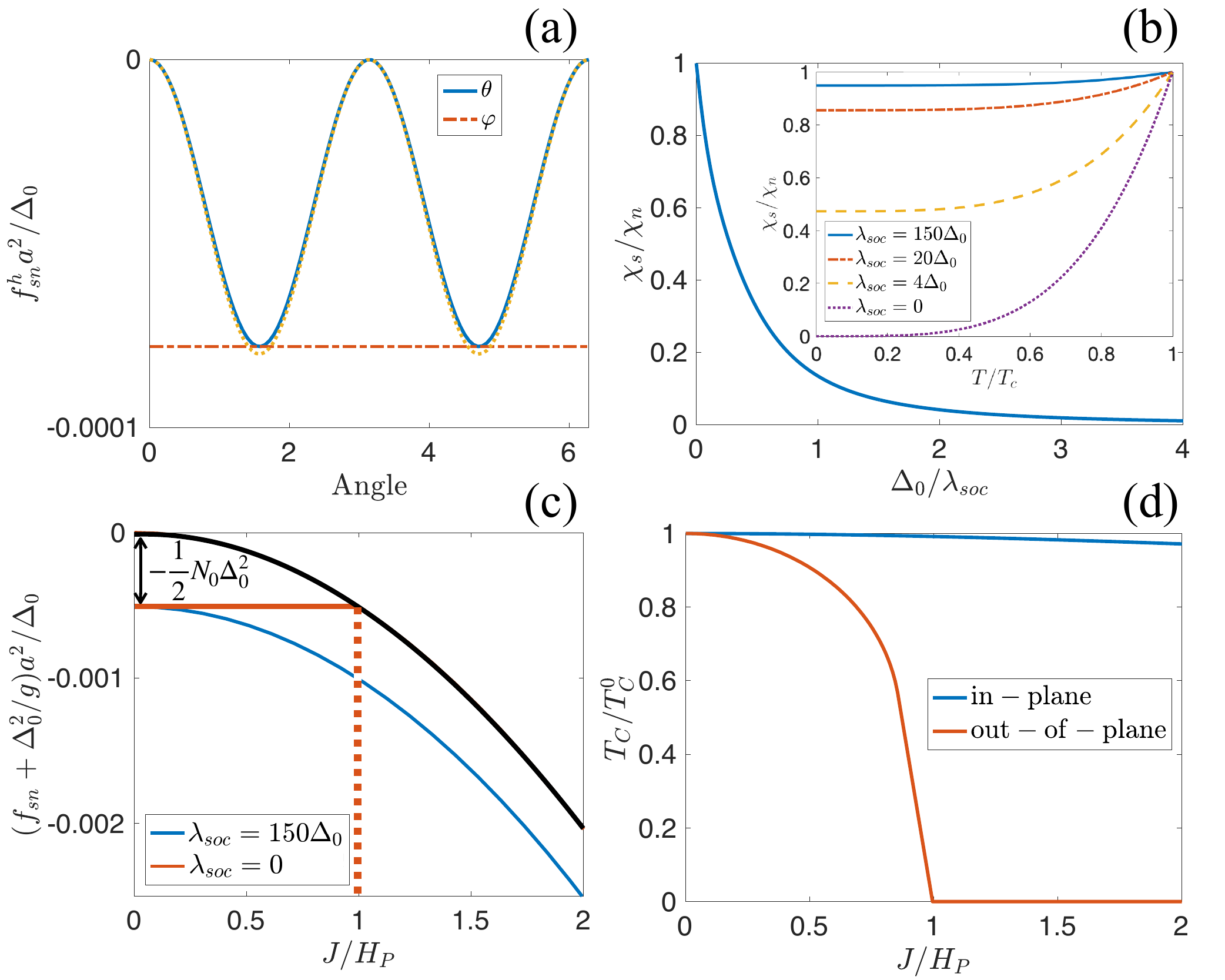}%
\caption{\label{fig:figure1} (a) Free energy density as a function of polar and azimuthal angles for the exchange field $J=0.5H_P$ and for the strength of the SOC $\lambda_{soc}=150\Delta_0$. Here $H_P=\Delta_0/\sqrt{2}$ is the Pauli paramagnetic limit, and $\Delta_0$ is the superconducting pair potential of NbSe$_2$ at $T=0$ and $J=0$. The yellow dashed line is the normal state energy $f_{sn}(\Delta=0)$. (b) Spin susceptibility at $T=0$ as a function of the superconducting pair potential $\Delta_0$. Inset: Spin susceptibility  as a function of temperature for various strengths of SOC $\lambda_{soc}$. Here $J=0.1H_P$. (c) Free energy density as a function of $J$ at $T=0$ in the presence and absence of the SOC. The black curve is the paramagnetic energy density $f_P$ in Eq.~\eqref{eq:ParamagneticEnergyDensity}.  (d) Critical temperature of in- and out-of-plane exchange fields as a function of exchange field strengths $J$, for  $\lambda_{soc}=150\Delta_0$. The critical temperature is normalized to the critical temperature at $J=0$.}
\end{figure}

Before the discussion of the magnetic vdW heterostructure, it is instructive to summarize the properties of Ising superconductivity in the proximity of one ferromagnetic layer. 
The free energy density difference between the superconducting and normal state energy $f_{sn}$ has exchange field independent and dependent parts. The exchange field dependent part $f_{sn}^h=f_{sn}(\boldsymbol{J})-f_{sn}(\boldsymbol{J}=0)$ is plotted as a function of the angles of the exchange field in Fig.~\ref{fig:figure1}(a) for $T=0$. We can see that the superconductor energetically prefers an in-plane exchange field ($\theta=\pi/2$). This is a direct consequence of Ising superconductivity. For conventional superconductors, the free energy density $f_{sn}$ is the same for both in-plane and out-of-plane exchange fields below the Pauli paramagnetic limit. In superconductors with a lack of inversion symmetry (such as NbSe$_2$), the Ising spin-orbit interaction gives rise to a non-zero and anisotropic spin susceptibility even at  $T=0$~\cite{frigeri2004spin,sigrist2009introduction,sohn2018unusual}. 

The free energy density $f_{sn}$ of an Ising superconductor in the presence of an exchange field at $T=0$ can be written as\footnote{The definition of spin susceptibility and related derivations are given in
appendix.~\ref{sec:AnalyEvaluFsn}}
\begin{equation}\label{eq:SingleLayerFreeEnergyDensitySmallh}
f_{sn}=-\frac{1}{2}N(0)\Delta_0^2-\frac{1}{2}\chi_{s}(T=0)J^2\sin^2\theta,
\end{equation}
where $N(0)$ is the density of states at the Fermi level, $\Delta_0$ is the superconducting pair potential at $T=0$ and $J=0$, and $\chi_{s}$ is the spin susceptibility for an in-plane exchange field describing the anisotropic spin response. For conventional superconductors, there is no such anisotropy, and the spin susceptibility vanishes at $T=0$ in the absence of spin relaxation.~\cite{PhysRev.110.769}. 

The amount of anisotropy depends on the size of the spin-orbit coupling $\lambda_{\rm soc}$ so that for very large $\lambda_{\rm soc}$, the spin susceptibility $\chi_s$ for in-plane exchange field approaches the normal-state spin susceptibility. This is shown in Fig.~\ref{fig:figure1}(b) as a function of $\Delta_0/\lambda_{\rm soc}$ at $T=0$ and in the inset as a function of temperature. On the other hand, for an exchange field perpendicular to the plane, the spin susceptibility corresponds to the case with $\lambda_{\rm soc}=0$.

The anisotropic spin susceptibility in Ising superconductors also leads to an
enhancement of the paramagetic critical field~\cite{noncentro2012,youn2012role}. In the presence of an exchange field, the normal-state free energy density is lowered by the term $-\frac{1}{2}\chi_n(\theta)J^2$. When the exchange field breaks the superconducting order, $f_{sn}$ equals the paramagnetic energy density $f_P$ in the normal state 
\begin{equation}\label{eq:ParamagneticEnergyDensity}
f_{P}=-\frac{1}{2}\chi_n(\theta)J^2,
\end{equation}
where $\chi_n(\theta)$ is the spin suceptibility in the normal state. Equating $f_{sn}$ and $f_P$  yields
$
J_c=\sqrt{\frac{N(0)}{\chi_n(\theta)-\chi_s(T=0)\sin^2\theta}}\Delta_0.
$
Here  $\chi_n(\theta)=\chi_P\cos^2\theta + \chi_n^{\rm soc} \sin^2\theta$, where $\chi_P=2N_0$ is the Pauli paramagnetic susceptibility, and $\chi_n^{\rm soc}$ is the normal state spin susceptibility caused by Ising SOC. For an out-of-plane field ($\theta=0$), the spin susceptibility is $\chi_P$ and the superconductivity experiences a first-order transition to the normal state, as shown in the red curve in Fig.~\ref{fig:figure1}(c). For $\theta=\pi/2$, the susceptibility is completely determined by $\chi_n^{\rm soc}$ and there is a second-order transition to the normal state at $J_c\gg H_P$. For the case of monolayer TMDs, $\chi_n^{\rm soc}\gg \chi_P$. For weak SOC, $\chi_n^{\rm soc}\rightarrow \chi_P$, and the normal state susceptibility is isotropic~\cite{frigeri2004spin}. The critical temperature is also influenced by this anisotropic spin susceptibility. In Fig.~\ref{fig:figure1}(d), the critical temperature $T_c$ is shown for in- and out-of-plane exchange fields as a function of the exchange field strength $J$. One can see that $T_c$ remains nearly unchanged for increasing in-plane exchange field, but the superconductivity is destroyed at the Pauli paramagnetic limit for the out-of-plane exchange field.

The strength of the induced exchange field in the superconductor depends on many factors~\cite{izyumov2002competition,RevModPhys.90.041001,CrCl32021}. If the induced exchange field is out-of-plane and is larger than the Pauli paramagnetic limit, $J>H_P$, the monolayer TMD remains in the normal state. However, the magnitude of $J$ depends on the chosen materials combination, and can also be below the critical field $H_P$. In what follows, we consider a case of the induced exchange field $J<H_P$ in the monolayer TMD~\cite{CrI32017,CrCl32021,doi:10.1126/sciadv.abb9379} so that it becomes superconducting at low enough temperature.

The direction of the exchange field is related to the magnetization direction of the ferromagnet, which is determined by the anisotropy energy. A monolayer ferromagnet such as ${\rm{CrI}_3}$~\cite{Huang2017} and CrBr$_3$~\cite{doi:10.1126/sciadv.abb9379}, shows
a uniaxial anisotropy, and the anisotropy energy density can be written as
\begin{equation}
U_{{\rm{aniso}}}=K_{{\rm{eff}}}\sin^2\theta,
\end{equation}
where $\theta$ is the polar angle of the magnetization direction with respect to the $z$ axis, and $K_{{\rm{eff}}}$ is the effective anisotropy constant with the unit of energy density. The main source of the effective anisotropy constant $K_{{\rm{eff}}}$ is the spin-orbit interaction\cite{lado2017origin,chen2020magnetic,bacaksiz2021distinctive}. The positive $K_{{\rm{eff}}}>0$ means the magnetization direction is along the easy-axis of the ferromagnet, and the negative $K_{{\rm{eff}}}<0$ would mean the magnetization direction is in the plane.

In the normal state, the anisotropic spin susceptibility caused by the spin-orbit interaction favors in-plane magnetic alignment, see Fig.~\ref{fig:figure1}(a). In the superconducting state, this tendency persists, but becomes weaker because the spin susceptibility is smaller in the superconducting state, as shown in Fig.~\ref{fig:figure1}(b).  Then the direction of the exchange field in the Ising  superconductor is determined by the competition of the anisotropy energy of the ferromagnet and the contribution of spin susceptibility to the free energy density of the superconductor. The part of total free energy density dependent on magnetization configuration can be written as

\begin{equation}
f_{{\rm{aniso}}}=\left(K_{\rm{eff}}-\frac{1}{2}\chi_SJ^2\right)\sin^2\theta. 
\end{equation}
If the condition $K_{\rm{eff}}<\chi_SJ^2/2$ holds, then the  direction of the exchange field is in the plane. This relation is possible as the anisotropy energy of a monolayer ferromagnet can be tuned by various means. For example, the magnetic anisotropy can be engineered by  strain~\cite{PhysRevB.98.144411} and by controlling the SOC via chemical tuning~\cite{doi:10.1126/sciadv.abb9379}. In other words, the anisotropic spin susceptibility leads to the switching of an off-plane easy axis anisotropy to the easy plane anisotropy if $K_{\rm{eff}}<\chi_SJ^2/2$. With this mechanism in mind, let us now discuss
how the interplay between superconductivity driven-magnetic switching would affect the coupling between two ferromagnets.

\section{Ising-mediated magnetic hysteresis}
\label{sec:main}

In the following, we consider a Ferromagnet/Superconductor/Ferromagnet (F/S/F) structure,
in which the easy axis of the ferromagnets is influenced by the presence of superconductivity. The magnetic moments in the ferromagnets are coupled through the superconducting layer, and this coupling also determines the direction of the induced exchange field. The full energetics of the system can be written as
\begin{equation}
f_{{\rm{aniso}}}=K_1\sin^2\theta_1+K_2\sin^2\theta_2+\delta f_{sn},
\end{equation}
where $K_{1/2}$ are the effective anisotropy constants of the two ferromagnets, $\theta_{1/2}$ are the related polar angles, and $\delta f_{sn}$ is the contribution of superconductivity to the anisotropy energy density. 

It is first worth noting how this mechanism would work
in a conventional superconductor. For a conventional superconductor with thickness smaller than the superconducting coherence length, the coupling between the ferromagnets through the superconducting layer gives rise to an antiparallel configuration of the magnetization directions~\cite{de1966coupling,hauser1969coupling,ojajarvi2021dynamics}. The contribution of this coupling to the energetics of the system $\delta f_{sn}$ is evaluated as 
$
\delta f_{sn}=N(0)\bar{J}^2\cos^2\left[\frac{1}{2}\left(\theta_1-\theta_2 \right)\right],
$
where $\bar{J}$ would be the strength of the induced exchange field in the metallic film. {Because of this term, the antiparallel alignment of the magnetization directions is favoured in the system.} 
However, and in stark contrast, for the case of a vdW heterostructure containing a two-dimensional Ising superconductor, the ferromagnetic layers couple through the spin susceptibility
\footnote{Direct coupling mediated by NbSe$_2$
is stronger than indirect superexchange between layers}. 
We now write the components of the exchange field as $J_x=J(\sin\theta_1\cos\psi_1+\sin\theta_2\cos\psi_2)$, $J_y=J(\sin\theta_1\sin\psi_1+\sin\theta_2\sin\psi_2)$, $J_z=J(\cos\theta_1+\cos\theta_2)$ in the Hamiltonian in Eq.~\eqref{eq:ExchnageHamiltonian}, diagonalize the Hamiltonian in Eq.~\eqref{eq:TotalHamiltonian}, obtaining the contribution of this coupling to the energetics of the system $
\delta f_{sn}=-\frac{1}{2}\chi_sJ^2\left( \sin\theta_1+\sin\theta_2\right)^2.
$ The overall effect of this term is to lower the total anisotropy energy.  Hence a parallel alignment maximizes this effect. This is opposite to the case of a thin metallic superconductor discussed above, where an antiparallel alignment is energetically favoured. Note that, our discussion above neglects direct Heisenberg coupling between the magnetic monolayers. The nature of that coupling can be ferromagnetic or antiferromagnetic depending on the geometric details. 
A well known example is the case of CrI$_3$ bilayers\cite{Huang201722,Soriano2019,Sivadas2018,Ubrig2019}, where depending on the relative alignment between layers the coupling can be either ferromagnetic or antiferromagnetic. This contribution is a higher order superexchange depending on the valence states of each element, can be only captured properly with quantum chemistry density functional theory calculations and can potentially depend on the geometric alignment between monolayers\cite{Huang201722,Soriano2019,Sivadas2018,Ubrig2019}. 
This contribution would determine the relative orientation of the magnetization between the two magnetic monolayers, and for the sake of concreteness in Fig. \ref{fig:structure} was taken as ferromagnetic. 

The ferromagnetic coupling mediated by Ising superconductivity can be detected by a magnetic hysteresis measurement. We use the Stoner–Wohlfarth (SW) model~\cite{stoner1948mechanism,tannous2008stoner} to calculate the hysteresis loop. In the SW model, the directions of the magnetization density $M$ and the external applied field $H$ are characterized by different angles with respect to the easy axis (polar angles). In the system we are considering, there is also the contribution of the superconductor to the anisotropy energy. Then the magnetization density is subjected to the competition between the anisotropy energy and Zeeman energy caused by the external applied field. The total anisotropy energy density is written as
\begin{equation}
\begin{split}
&f_{{\rm{aniso}}}=K_1\sin^2\theta_1+K_2\sin^2\theta_2+\delta f_{sn}\\
&-HM_{s1}\cos\left(\theta_1-\vartheta \right)-HM_{s2}\cos\left(\theta_2-\vartheta \right),
\end{split}
\end{equation}
where $H$ is the external applied field, $M_{s1/2}$ is the saturation magnetization density in the two ferromagnets, and $\vartheta$ is the angle between the easy axis and the direction of the external magnetic field. {Substituting the value of $\delta f_{sn}$, we have
\begin{equation}\label{eq:HysAnisoEnergy}
\begin{split}
f_{{\rm{aniso}}}&=\left(K_1-\frac{1}{2}\chi_sJ^2\right)\sin^2\theta_1\\
&+\left(K_2-\frac{1}{2}\chi_sJ^2\right)\sin^2\theta_2-\chi_sJ^2\sin\theta_1\sin\theta_2\\
&-HM_{s1}\cos\left(\theta_1-\vartheta \right)-HM_{s2}\cos\left(\theta_2-\vartheta \right).
\end{split}
\end{equation}}The hysteresis curve is then obtained by calculating the longitudinal and transverse components of the magnetization density
\begin{equation}
M_z=M_{s1}\cos\theta_1^*+M_{s2}\cos\theta_2^*,
\end{equation}
\begin{equation}
M_x=M_{s1}\sin\theta_1^*+M_{s2}\sin\theta_2^*,
\end{equation}
where $\theta_{1/2}^*$ are the angles for which the total anisotropy energy density $f_{{\rm{aniso}}}$ is minimized. 

\begin{figure}[t]
\includegraphics[width=\linewidth]{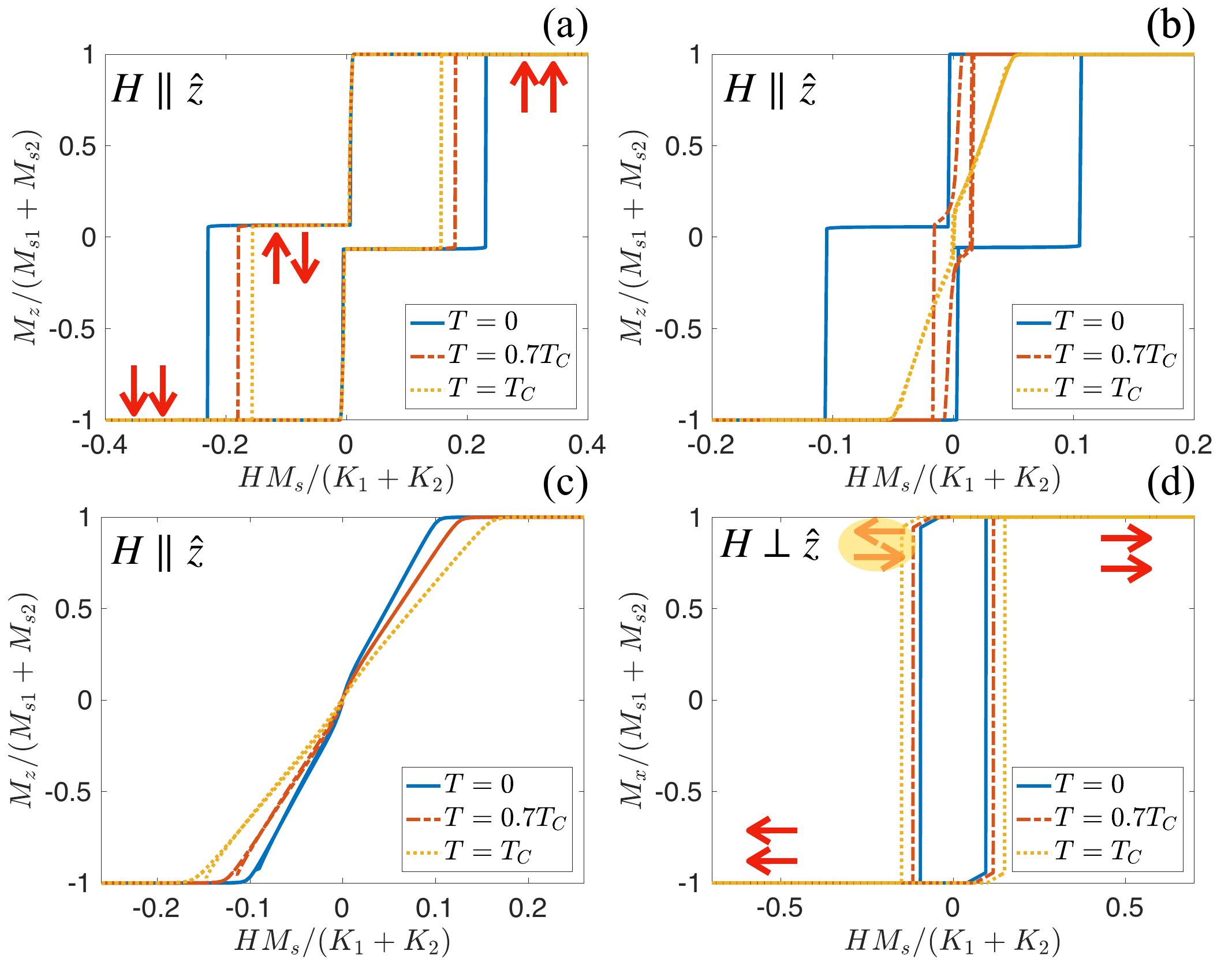}%
\caption{\label{fig:hysteresis} (a) Hysteresis curve for $K_1=1.05\chi_s(T=0)J^2$ for an out-of-plane external field. (b) Hysteresis curve for $K_1=0.95\chi_s(T=0)J^2$ for an out-of-plane external field. (c) Hysteresis curve for $K_1=0.85\chi_s(T=0)J^2$ for an out-of-plane external field. (d) Same with (c) but for in-plane external field. }
\end{figure}

For an easy axis ferromagnet such as CrBr$_3$ or CrI$_3$, the induced exchange field is out-of-plane. In the absence of the Ising superconductor, the magnetic alignment of the system is ferromagnetic and two coercive fields appear in the hysteresis curve at $H_{c1}M_{s}=\pm2K_1$ and $H_{c2}M_{s}=\pm2K_2$. If the anisotropy energy of the ferromagnet is larger than the contribution $\delta f_{sn}$ of the superconductor, the net effect of the Ising superconductor leads to a modification of the anisotropy constants, as can be seen from Eq.~\eqref{eq:HysAnisoEnergy}. The modification shifts the coercive fields as $H_{c1}M_{s}=\pm2(K_1-\chi_sJ^2)$ and $H_{c2}M_{s}=\pm2(K_2-\chi_sJ^2)$. We can see that, if the values of $K_1$ and $K_2$ are comparable with $\chi_sJ^2$, the temperature dependence of the anisotropic spin susceptibility causes significant modifications to the hysteresis curves. In order to see such modifications, the difference between $K_1$ and $K_2$ must be smaller than the difference of the spin susceptibility $\chi_s$ at $T_c$ and $T=0$. In the following calculations, we set $\lambda_{\rm soc}=150\Delta_0$, $K_1=1.12K_2$ and $M_{s1}=1.12M_{s2}=1.12M_s$ to illustrate how superconductivity affects the magnetic hysteresis in a temperature dependent manner. The hysteresis curve is shown in Fig.~\ref{fig:hysteresis}. It is worth noting that, while $\lambda_{\rm soc}/\Delta$ cannot be efficiently tuned in a specific material, specific ratios can be obtained by taking different dichalcogenides including 1H-TaS$_2$, 1H-TaSe$_2$, 1H-NbS$_2$, 1H-NbSe$_2$, and
potentially continuously using dichalcogenide alloys\cite{Zhao2019}.

The modified coercive fields for strong anisotropy fields $K_i>\chi_sJ^2$ are shown in the hysteresis curve in Fig.~\ref{fig:hysteresis}(a). Decreasing the temperature, $\chi_s$ becomes smaller, and the difference in the coercive fields becomes larger. In the opposite case $K_i<\chi_sJ^2$, the induced field is in the plane, and there is no hysteresis for an out-of-plane component of the magnetization density $M_z$, as shown in Fig.~\ref{fig:hysteresis}(c). The hysteresis curve for $M_x$ is shown in Fig.~\ref{fig:hysteresis}(d).  Contrary to the case in Fig.~\ref{fig:hysteresis}(a), the location of the coercive fields moves closer at lower temperature for smaller $\chi_s$. 

A more interesting case takes place for ferromagnets with anisotropy constants with intermediate values\cite{Tartaglia2020}. In this case, the induced exchange field is out-of-plane at low temperature and in-plane at a higher temperature. This transition takes place at a temperature $T=T^*$ and is due to the temperature dependence of the spin susceptibility $\chi_s$, so that the induced field is in-plane for a temperature $T>T^*$. This case is shown in Fig.~\ref{fig:hysteresis}(b). 
In this limit, an in-plane exchange field is favoured in Ising superconducting and normal states for $T>T^*$. Note that, besides moving the coercive fields of single magnets, NbSe$_2$ promotes a ferromagnetic coupling between the magnets in the normal state. This affects the hysteresis curves by reducing the region with antiparallel magnetization directions, but cannot be directly measured in situ because of a lack of a control measurement without the presence of NbSe$_2$. In the superconducting state this ferromagnetic coupling decreases but, unlike the coupling mediated by a conventional superconductor, does not become antiferromagnetic.

It is also worth noting that, beyond the NbSe$_2$ platform considered here, this phenomenology
can also potentially take place for generic monolayer Ising superconductors, for example electron doped MoS$_2$~\cite{lu2015evidence,costanzo2016gate} and other TMD monolayers like TaS$_2$/TaSe$_2$~\cite{PrintableLi2020}. For other potential candidates with smaller spin-orbit coupling, the dependence of the anisotropic spin susceptibility on temperature is stronger, see Fig.~\ref{fig:figure1}(b). Then the hysteresis curves can be obtained for more varied anisotropy fields.

Interestingly, while magnetic order is often controlled by external knobs such as magnetic field, here we propose a radically different mechanism in which the magnetic order itself is controlled by the superconducting order  in one of the components of the heterostructure.
These results highlight that Ising superconductivity controls the way how the system reacts to magnetic fields, an effect that can be present in a variety of superconductor-ferromagnet van der Waals heterostructures. From the practical point of view, we exploit the unique temperature dependence of $\chi_s$, to uncover the interplay between the magnetism and superconductivity.

Besides magnetic hysteresis that in 2D materials can be measured with the magneto-optical Kerr effect, the anisotropic spin susceptibility of the Ising superconductivity can be accessed also via the ferromagnetic resonance of the multilayer system.\cite{ojajarvi2021dynamics}

For the possible experimental realization, below we briefly elaborate on the most promising materials
candidates based on recent experiments.
First, recent experiments demonstrated heterostructures of the monolayer ferromagnet with CrBr$_3$
on bulk NbSe$_2$~\cite{Kezilebieke2020}, making CrBr$_3$ a promising candidate for our proposal.
While the experiment focused on a monolayer CrBr$_3$ on top of 
a bulk NbSe$_2$, it is worth noting that the value of the
exchange proximity would be comparable for a monolayer NbSe$_2$ given
its van der Waals nature.
From the point of view of the ferromagnetic insulator candidates,
apart from CrBr$_3$,
CrI$_3$~\cite{CrI32017}, CrCl$_3$~\cite{CrCl32021} and recently
synthetised chromium heterohalides~\cite{doi:10.1126/sciadv.abb9379} would provide excellent candidates. Importantly, van der Waals chromium heterohalides have been demonstrated to have
a completely tunable magnetic anisotropy, making them
outstanding candidates for our proposal~\cite{doi:10.1126/sciadv.abb9379}. Finally, we note that the magnetic anisotropy could also be engineered by strain~\cite{PhysRevB.98.144411}, yet this approach could be more challenging
from the experimental point of view.

It is finally worth mentioning that the possible moir\'{e} patterns between the layers~\cite{magneticmoire2019,wang2020correlated} and correlation effects~\cite{HUANG2016997,wang2020correlated} in the TMDs have an influence on the effective values of the parameters used in our model. These effects modify the band structure of the middle layer, especially the splitting around the K points. However, the correlation effects in the ferromagnetic layers, for example magnetic excitations~\cite{PhysRevX.8.041028,costa2020topological,jin2018raman} do not directly influence the results, as these effects do not modify the form of the low energy Hamiltonian for the hybrid heterostructure we are considering.

\section{Conclusions}
\label{sec:con}
To summarize, we demonstrate how Ising superconductivity allows controlling the magnetic coupling
in ferromagnet/NbSe$_2$/ferromagnet heterostructures.
In particular, we show that the anisotropic spin response
inherited by the Ising superconductor
allows flipping the magnetic alignment upon entering the
superconducting state.
We show that the contribution of the Ising superconductivity to the energetics of the system stems from an anisotropic spin susceptibility. In the magnetic vdW heterostructure, the coupling between the
ferromagnets is achieved through the anisotropic spin susceptibility, which gives rise to a parallel alignment of the
magnetization directions, both in the superconducting and normal states. 
Interestingly, such magnetic switching
was shown to lead to a hysteretic behavior in the magnetic alignment
purely driven by the spin susceptibility of the NbSe$_2$
inherited from Ising spin-orbit coupling.
Our results put forward superconductor/ferromagnet van der Waals
heterostructures as a novel platform to explore
superconducting spintronics phenomena dominated
by Ising spin-orbit coupling effects.

\textbf{Acknowledgments:}
We acknowledge
the computational resources provided by
the Aalto Science-IT project,
and the financial support from the
Academy of Finland Projects No.
331342, No. 336243 and No. 317118, 
and the Jane and Aatos Erkko Foundation.

\appendix

\section{Single-Band Tight-Binding Model\label{sec:EffectiveTBModel}}

Here we elaborate on the low energy model used for NbSe$_2$. We take a single
band model in a triangular lattice as shown in
Eq.~\eqref{eq:OneBandTighBindingHamiltonian},
whose Bloch Hamiltonian takes the form

\begin{equation}\label{eq:MEETBHamiltonian1}
\begin{split}
&\xi_{\boldsymbol{k}\uparrow}=2t_0\left[\cos\left(2\alpha-\phi_0 \right)+2\cos\left(\beta \right)\cos\left(\alpha+\phi_0 \right) \right]\\
&+2t_1\left[\cos\left(3\alpha-\beta-\phi_1 \right)+\cos\left(2\beta-\phi_1 \right)\right.\\
&\left.+\cos\left(3\alpha+\beta+\phi_1 \right) \right]+2t_2\left[\cos\left(2\alpha+2\beta-\phi_2 \right)\right.\\
&\left.+\cos\left(2\alpha-2\beta-\phi_2 \right)+\cos\left(4\alpha+\phi_2 \right) \right]\\
&+2t_3\left[\cos\left(5\alpha+\beta-\phi_3 \right)+\cos\left(4\alpha-2\beta-\phi_3 \right)\right.\\
&\left.+\cos\left(4\alpha+2\beta+\phi_3 \right)+\cos\left(\alpha-3\beta+\phi_3 \right) \right. \\
&+\left. \cos\left(5\alpha-\beta+\phi_3 \right)+\cos\left(\alpha+3\beta+\phi_3 \right) \right]\\ &+2t_4\left[\cos\left(3\alpha-3\beta-\phi_4 \right)+\cos\left(3\alpha+3\beta-\phi_4 \right)\right.\\
&\left.+\cos\left(6\alpha+\phi_4 \right) \right]+2t_5\left[\cos\left(2\alpha-\beta-\phi_5 \right)\right.\\
&\left.+\cos\left(4\beta-\phi_5 \right)+\cos\left(6\alpha+2\beta+\phi_5 \right) \right]+\varepsilon
\end{split}
\end{equation}
\begin{equation}\label{eq:MEETBHamiltonian2}
\begin{split}
&\xi_{\boldsymbol{k}\downarrow}=2t_0\left[\cos\left(2\alpha+\phi_0 \right)+2\cos\left(\beta \right)\cos\left(\alpha-\phi_0 \right) \right]\\
&+2t_1\left[\cos\left(3\alpha-\beta+\phi_1 \right)+\cos\left(2\beta+\phi_1 \right)\right.\\
&\left.+\cos\left(3\alpha+\beta-\phi_1 \right) \right]+2t_2\left[\cos\left(2\alpha+2\beta+\phi_2 \right)\right.\\
&\left.+\cos\left(2\alpha-2\beta+\phi_2 \right)+\cos\left(4\alpha-\phi_2 \right) \right]\\
&+2t_3\left[\cos\left(5\alpha+\beta+\phi_3 \right)+\cos\left(4\alpha-2\beta+\phi_3 \right)\right.\\
&\left.+\cos\left(4\alpha+2\beta-\phi_3 \right)+\cos\left(\alpha-3\beta-\phi_3 \right) \right.\\
&+\left. \cos\left(5\alpha-\beta-\phi_3 \right)+\cos\left(\alpha+3\beta-\phi_3 \right) \right]\\ &+2t_4\left[\cos\left(3\alpha-3\beta+\phi_4 \right)+\cos\left(3\alpha+3\beta+\phi_4 \right)\right.\\
&\left.+\cos\left(6\alpha-\phi_4 \right) \right]+2t_5\left[\cos\left(2\alpha-\beta+\phi_5 \right)\right.\\
&\left.+\cos\left(4\beta+\phi_5 \right)+\cos\left(6\alpha+2\beta-\phi_5 \right) \right]+\varepsilon,
\end{split}
\end{equation}
where $t_i$s are the hopping energies between first to sixth neighbours, $\phi_i$ are corresponding phases,  $\epsilon$ is the on-site energy, $\alpha=k_xa/2$, $\beta=\sqrt{3}k_ya/2$, and $a$ is the lattice constant shown in Fig.~\ref{fig:tight_binding}(a). 

The tight-binding parameters are determined by comparing the effective one-band model with the three-band model in Ref.~\onlinecite{he2018magnetic}, and listed in Table.~\ref{tab:effective_Hamiltonian_parameters}.

\begin{table}[h]
\caption{\label{tab:effective_Hamiltonian_parameters} Parameters of the one-band tight-binding model fitted to the three band model.}
\begin{ruledtabular}
\begin{tabular}{cccccccccccc}
$\varepsilon$ & $t_0$ & $t_1$ & $t_2$ & $t_{3}$ \\ 
-44.5 meV & 26.3 meV  & 99.1 meV & -1.4 meV & -11.2 meV \\
\hline
$t_{4}$ & $t_{5}$ & $\phi_0$ & $\phi_{1\sim5}$\\
-14.6 meV & 2.5 meV & 0.6 & 0
\end{tabular}
\end{ruledtabular}
\end{table}

The momentum dependent splitting can be obtained from the difference between the matrix component of the Hamiltonian in Eq.~\eqref{eq:MEETBHamiltonian1} and Eq.~\eqref{eq:MEETBHamiltonian2}
\begin{equation}\label{eq:StrenthmSOC}
\Delta_{soc}=\xi_{\boldsymbol{k}\uparrow}-\xi_{\boldsymbol{k}\downarrow}=8t_0\sin\phi_0\left(\cos\alpha-\cos\beta \right)\sin\alpha.
\end{equation}
The strength of the SOC can be defined as the half of the splitting at a $K$ (or $K'$) point
\begin{equation}
\lambda_{soc}=\frac{1}{2}\vert\Delta_{soc}(K)\vert=3\sqrt{3}t_0\sin\phi_0.
\end{equation}
For monolayer NbSe$_2$, the parameters in Table \ref{tab:effective_Hamiltonian_parameters} give $\lambda_{soc}=77.2\ \rm{meV}$. Note that the tight-binding parameters are fitted first in the absence of SOC ($\phi_i=0$), and $\phi_i$s are determined in the presence of SOC with respect to the splitting between energy bands. As the effective value of $\lambda_{soc}$ is influenced by the presence of charge density wave order\cite{PhysRevB.97.081101}, we consider different $\lambda_{soc}$ by varying $\phi_0$.

\section{Self-Consistency Equation}
\label{sec:SelConEqn}

The self-consistency equation can be derived from the free energy density by minimizing with respect to the superconducting pair potential. Using the free energy density expression in Eq.~\eqref{eq:FreeEnergyDensityGeneral}, we have
\begin{equation}\label{eq:MinimizingFreeEnergyDensity0}
\frac{\partial f_{sn}}{\partial \Delta}=\frac{2\Delta}{g}-\sum_{\boldsymbol{k},n}\frac{\partial\epsilon_{\boldsymbol{k},n}}{\partial \Delta}\tanh\left(\frac{\epsilon_{\boldsymbol{k},n}}{2k_BT} \right)=0,
\end{equation}
where $\epsilon_{\boldsymbol{k},n}$ is the eigenvalue of the Hamiltonian in Eq.~\eqref{eq:TotalHamiltonian}. 
For convenience, we now write the partial derivative of $\epsilon_{\boldsymbol{k},n}$ with respect to $\Delta$ in terms of $\epsilon_{\boldsymbol{k},n}$ as
\begin{equation}
\begin{split}
&\frac{\partial\epsilon_{\boldsymbol{k},n}}{\partial \Delta}=\\
&\frac{\Delta\left(-2\epsilon_{\boldsymbol{k},n}^2+\xi_{\boldsymbol{k}\downarrow}^2+\xi_{\boldsymbol{k}\uparrow}^2+2\Delta^2-2J^2 \right)}{\left(\xi_{\boldsymbol{k}\uparrow}^2-\xi_{\boldsymbol{k}\downarrow}^2 \right)h_z+\epsilon_{\boldsymbol{k},n}\left( \xi_{\boldsymbol{k}\downarrow}^2+\xi_{\boldsymbol{k}\uparrow}^2+2\Delta^2+2J^2 \right)-2\epsilon_{\boldsymbol{k},n}^3}. 
\end{split}
\end{equation}
Changing the momentum sum to a momentum integral, we can write the self-consistency equation as
\begin{equation}
\begin{split}
&\frac{g}{2}\int_{BZ}\frac{d\boldsymbol{k}}{(2\pi)^2}\left[ \sum_{n} \tanh\left(\frac{\epsilon_{\boldsymbol{k},n}}{2k_BT}
\right)\times\right.\\
&\left. \frac{\xi_{\boldsymbol{k}+}^2-J^2+\Delta^2 -\epsilon_{\boldsymbol{k},n}^2}{\xi_{\boldsymbol{k}-}^2h\cos\theta+\epsilon_{\boldsymbol{k},n}\left(\xi_{\boldsymbol{k}+}^2+J^2+\Delta^2 -\epsilon_{\boldsymbol{k},n}^2\right)}\right]=1,
\end{split}
\end{equation}
where $\xi_{\boldsymbol{k}+}^2=(\xi_{\boldsymbol{k}\downarrow}^2+\xi_{\boldsymbol{k}\uparrow}^2)/2$ and $\xi_{\boldsymbol{k}-}^2=(\xi_{\boldsymbol{k}\uparrow}^2-\xi_{\boldsymbol{k}\downarrow}^2)/2$. The interaction potential $g$ could be determined from the superconducting pair potential of monolayer $\rm{NbSe}_2$ at zero temperature and exchange field, $\Delta_0=1.8T_{c0}$, where $T_{c0}=3$ K~\cite{xi2016ising,sergio2018tuning}. The momentum integral goes over the first Brillouin zone. This model hence disregards the possible retardation effects in the source of the attractive interaction. For the value of $\Delta$ small compared to the bandwidth considered in this manuscript, this simplification only affects the chosen value of the coupling constant $g$ with which $\Delta$ is obtained.

\section{Free Energy Density at Zero Temperature}
\label{sec:AnalyEvaluFsn}

The Hamiltonian in Eq.~\eqref{eq:HamiltonianSCi} can be written in the matrix form as
\begin{equation}\label{eq:BdGHamiltonianmarix}
H_{{\rm{BdG}}}=\begin{pmatrix}
\hat{H}_0(\boldsymbol{k})-\boldsymbol{J}\cdot\boldsymbol{\sigma} & \hat{\Delta}\\
\hat{\Delta}^{\dagger} & -\sigma_y\left[\hat{H}_0^*(-\boldsymbol{k})-\boldsymbol{J}\cdot\boldsymbol{\sigma}\right]\sigma_y
\end{pmatrix},
\end{equation}
where $\hat{H}_0(\boldsymbol{k})={\rm{diag}}(\xi_{\boldsymbol{k}\uparrow},\xi_{\boldsymbol{k}\downarrow})$,  $\xi_{\boldsymbol{k}\uparrow\downarrow}$ is given in Eq.~\eqref{eq:MEETBHamiltonian1} and Eq.~\eqref{eq:MEETBHamiltonian2}, $\hat{\Delta}=\Delta\tau_1$, $\boldsymbol{J}$ is the induced exchange field in the superconductor, and $\sigma_i/\tau_i$ is the Pauli matrix in the spin/Nambu space. Up to the second order in $J=\vert\boldsymbol{J}\vert$, the eigenvalues of the Hamiltonian in Eq.~\eqref{eq:BdGHamiltonianmarix} can be written as
\begin{equation}\label{eq:AnalyEigenvalues1}
\begin{split}
E_{\boldsymbol{k}\uparrow}=&\pm\sqrt{\xi_{\boldsymbol{k}\uparrow}^2+\Delta_0^2}+J\cos\theta\\
&\pm \frac{(\xi_{\boldsymbol{k}\uparrow}\xi_{\boldsymbol{k}\downarrow}+\xi_{\boldsymbol{k}\uparrow}^2+2\Delta_0^2)}{(\xi_{\boldsymbol{k}\uparrow}^2-\xi_{\boldsymbol{k}\downarrow}^2)\sqrt{\xi_{\boldsymbol{k}\uparrow}^2+\Delta_0^2}}J^2\sin^2\theta
\end{split}
\end{equation}
\begin{equation}\label{eq:AnalyEigenvalues2}
\begin{split}
E_{\boldsymbol{k}\downarrow}=&\pm\sqrt{\xi_{\boldsymbol{k}\downarrow}^2+\Delta_0^2}-J\cos\theta\\
&\pm \frac{(\xi_{\boldsymbol{k}\uparrow}\xi_{\boldsymbol{k}\downarrow}+\xi_{\boldsymbol{k}\downarrow}^2+2\Delta_0^2)}{(\xi_{\boldsymbol{k}\downarrow}^2-\xi_{\boldsymbol{k}\uparrow}^2)\sqrt{\xi_{\boldsymbol{k}\downarrow}^2+\Delta_0^2}}J^2\sin^2\theta,
\end{split}
\end{equation}
where $\theta$ is the polar angle of the exchange field.
We note that in the presence of an in-plane component of the exchange field,
$\uparrow,\downarrow$ denote the pseudospin degree of freedom,
instead of the original $z$ spin component.

At $T=0$, the free energy density can be simplified as
\begin{equation}\label{eq:FreeEnergyDensityAtZeroTemperature}
f(T=0)=\frac{\Delta_0^2}{g}-\sum_{\boldsymbol{k},n}\epsilon_{\boldsymbol{k},n}.
\end{equation}
Then we have 
\begin{equation}\label{eq:AnalyFsnIntegral}
\begin{split}
f_{sn}=&-\int\frac{d\boldsymbol{k}}{(2\pi)^2}\left(\sqrt{\xi_{\boldsymbol{k}\uparrow}^2+\Delta_0^2}-\xi_{\boldsymbol{k}\uparrow}+\sqrt{\xi_{\boldsymbol{k}\downarrow}^2+\Delta_0^2}-\xi_{\boldsymbol{k}\downarrow}\right)\\
&-\int\frac{d\boldsymbol{k}}{(2\pi)^2}\left[\frac{(\xi_{\boldsymbol{k}\uparrow}\xi_{\boldsymbol{k}\downarrow}+\xi_{\boldsymbol{k}\uparrow}^2+2\Delta_0^2)}{(\xi_{\boldsymbol{k}\uparrow}^2-\xi_{\boldsymbol{k}\downarrow}^2)\sqrt{\xi_{\boldsymbol{k}\uparrow}^2+\Delta_0^2}}\right.\\
&\left.+\frac{(\xi_{\boldsymbol{k}\uparrow}\xi_{\boldsymbol{k}\downarrow}+\xi_{\boldsymbol{k}\downarrow}^2+2\Delta_0^2)}{(\xi_{\boldsymbol{k}\downarrow}^2-\xi_{\boldsymbol{k}\uparrow}^2)\sqrt{\xi_{\boldsymbol{k}\downarrow}^2+\Delta_0^2}} \right]J^2\sin^2\theta+\frac{\Delta_0^2}{g}.
\end{split}
\end{equation}

The integrals in the first row can be solved by changing the momentum integral to an energy integral, which yields $-N(0)\Delta^2/2-\Delta^2/g$, where $N(0)$ is the density of states at the Fermi level, and $g$ is the interaction potential. The integral in the second row is related to the spin susceptibility due to the definition $\chi_{\rm{spin}}=-\partial^2 f_{sn}/\partial J^2\vert_{J\rightarrow0}$. From Eq.~\eqref{eq:FreeEnergyDensityGeneral}, we have
\begin{equation}
\begin{split}
\chi_{i}=&-\frac{\partial^2 f}{\partial J_{i}^2}\Big\vert_{J_{i}\rightarrow0}=\sum_{\boldsymbol{k},n}\left[\frac{1}{2T}{\rm{sech}}^2\left(\frac{\epsilon_{\boldsymbol{k},n}}{2T} \right)\left(\frac{\partial \epsilon_{\boldsymbol{k},n}}{\partial J_{i}} \right)^2\right.\\
&\left.+\tanh\left(\frac{\epsilon_{\boldsymbol{k},n}}{2T}\right)\left(\frac{\partial^2 \epsilon_{\boldsymbol{k},n}}{\partial J_{i}^2} \right) \right]\Bigg\vert_{J_{i}\rightarrow0},
\end{split}
\end{equation}
where $i=\parallel/\perp$. At $T=0$, from the eigenvalues in Eq.~\eqref{eq:AnalyEigenvalues1} and Eq.~\eqref{eq:AnalyEigenvalues2}, we have $\chi_{\perp}=0$, and the second integral in Eq.~\eqref{eq:AnalyFsnIntegral} is equal to the spin susceptibility $\chi_{\parallel}=\chi_s$. Now we can write
\begin{equation}\label{eq:PTTotalFreeEnergyDensityFull}
f_{sn}=-\frac{1}{2}N(0)\Delta_0^2-\frac{1}{2}\chi_sJ^2\sin^2\theta.
\end{equation}
Contrary to conventional superconductors with weak spin-orbit coupling, the spin susceptibility is nonzero at zero temperature. It leads to several important properties of Ising superconductivity like enhancing the critical field of a superconductor in the presence of an in-plane exchange field. 

The spin susceptibility can be solved analytically for the case of weak SOC. For such a case, the spin-splitting caused by SOC can be approximated by the splitting at the Fermi level. We can write $\xi_{\boldsymbol{k}\uparrow}=\xi_{\boldsymbol{k}}+\lambda_0$ and $\xi_{\boldsymbol{k}\downarrow}=\xi_{\boldsymbol{k}}-\lambda_0$, where $\lambda_0=\vert\Delta_{soc}(k_F)\vert/2$, and $\Delta_{soc}$ is defined in Eq.~\eqref{eq:StrenthmSOC}. Then the momentum integral can be transformed to energy integral, and the spin susceptibility is given by
\begin{equation}
\chi_s=\chi_P\left[1-\frac{\Delta_0^2}{2\lambda_0\sqrt{\lambda_0^2+\Delta_0^2}}\log\left(\frac{\lambda_0+\sqrt{\lambda_0^2+\Delta^2}}{-\lambda_0+\sqrt{\lambda_0^2+\Delta^2}} \right) \right],
\end{equation}
where $\chi_P=2N_0$ is the Pauli paramagnetic susceptibility. We can see that for weak SOC, $\chi_n^{soc}\rightarrow\chi_P$, and the structure of the spin susceptibility is consistent with
known results~\cite{frigeri2004spin,gor2001superconducting}. For strong SOC, namely, $\lambda_{soc}\sim t_0$, the spin susceptibility cannot be solved analytically and the numerically calculated results are shown in the main text.

\bibliography{apssamp}

\end{document}